# Complete Suppression of Phase Segregation in Mixed-Halide Perovskite Nanocrystals under Periodic Heating

Shengnan Feng[1†], Rentong Duan[1†], Yu Ju[1], Shuyi Li[1], Chunfeng Zhang[1], Shuxia Tao[2*], Min Xiao[1,3*], and Xiaoyong Wang[1*]

[1]*National Laboratory of Solid State Microstructures, School of Physics, and Collaborative Innovation Center of Advanced Microstructures, Nanjing University, Nanjing 210093, China*

[2]*Materials Simulation and Modelling & Center for Computational Energy Research, Department of Applied Physics, Eindhoven University of Technology, 5600 MB Eindhoven, The Netherlands*

[3]*Department of Physics, University of Arkansas, Fayetteville, Arkansas 72701, USA*

*Correspondence to S.T. (S.X.Tao@Tue.nl), M.X. (mxiao@uark.edu) or X.W. (wxiaoyong@nju.edu.cn)

†These authors contributed equally to this work

**Under continuous light illumination, it is known that localized domains with segregated halide compositions form in semiconducting mixed-halide perovskites, thus severely limiting their optoelectronic applications due to the negative changes in bandgap energies and charge-carrier characteristics. Here we deposit mixed-halide perovskite $CsPbBr_{1.2}I_{1.8}$ nanocrystals onto an indium tin oxide substrate, whose temperature can be rapidly changed by ~10 ºC in a few seconds by applying or removing an external voltage. Such a sudden temperature change induces a temporary transition of $CsPbBr_{1.2}I_{1.8}$ nanocrystals from the segregated phase to the mixed phase, the latter of which can be permanently**

**maintained when the light illumination is coupled with periodic heating cycles. These findings mark the emergence of a practical solution to the detrimental phase-segregation problem, given that a small temperature modulation is readily available in various fundamental studies and practical devices using mixed-halide perovskites.**

Lead mixed-halide perovskites of $APbBr_xI_{3-x}$ ($0 < x < 3$), with A being $CH_3NH_3^+$ ($MA^+$), $HC(NH_2)_2^+$ ($FA^+$) or $Cs^+$, are featured with high fluorescence quantum yield[1], long carrier diffusion length[2], good defect tolerance[3], and tunable emission wavelength covering a wide spectral range from the visible to the near infrared[4,5]. These superior optoelectronic properties have rendered them excellent candidates as absorbers and emitters for highly-efficient solar cells[6,7] and light-emitting diodes[8,9]. However, the primary challenge for the further advancements is the instability issue, which is manifested as lowered open-circuit voltage[10] and decreased emission-color purity[11] under long-term operations. The instability of mixed-halide perovskites is closely linked to the phase-segregation effect[12-15], wherein light illumination or electrical biasing would give rise to iodide- and bromide-rich domains that severely modify the otherwise homogeneous energy landscape for charge carriers. Several driving forces have been proposed so far for anion migration[16-18] in this intriguing process, mainly including polaronic strain caused by hole localization[19,20], density gradient or electric field imposed by charge carriers[16,21,22], free-energy variation of the system[23-26], and iodide oxidation/repulsion promoted by photogenerated holes[27-29]. Various strategies to improve the materials quality have been developed to mitigate the phase-segregation effect, e.g., increasing the grain size[30], improving the film crystallinity[31], passivating the surface defect[32], and mixing the A-site cations[33]. However, these alternations in materials could pose a stringent limit on the available bandgap energies and even induce unwanted changes of carrier recombination and transport dynamics[34].

The phase-segregation effect can also be controlled by some external stimuli such as hydrostatic pressure[35,36] and light intensity[26,37,38]. In the former case, a compressive strain can be induced by the external pressure[39], which is capable of suppressing phase segregation by increasing the activation energy for anion migration[39,40]. In the latter case, a threshold light intensity is required to yield the segregated phase[26,37], which, however, can be reversed to the mixed phase at sufficiently high light intensities due to polaron formation across the length scale larger than the average separation of segregated domains[38]. Despite providing in-depth information on the underlying mechanism, it is impractical to apply the large compressive strain or high light intensity to a working device of mixed-halide perovskites.

Here, we report on the complete suppression of phase segregation in mixed-halide $CsPbBr_{1.2}I_{1.8}$ nanocrystals (NCs) by means of a mild periodic heating method. This is achieved by depositing a dense film of $CsPbBr_{1.2}I_{1.8}$ NCs on top of an ITO (indium tin oxide) substrate, which is capable of changing the sample temperature by ~10 ºC rapidly upon the application or removal of a bias voltage. When the phase segregation has already been induced in the NCs under light illumination, a sudden temperature increase would convert them completely to the mixed phase within just several seconds. Intriguingly, the light-induced segregated phase of heated $CsPbBr_{1.2}I_{1.8}$ NCs can also be reverted to the mixed phase within several seconds after a sudden decrease of the sample temperature. After either of the above heating and cooling operations, the mixed phase of $CsPbBr_{1.2}I_{1.8}$ NCs can only be kept temporarily and they would finally switch back to the segregated phase under elongated light illumination. However, when continuous light illumination is coupled with periodic heating durations realized by alternatively switching on and off a bias voltage applied to the ITO substrate, the $CsPbBr_{1.2}I_{1.8}$

NCs would remain at the mixed phase without any detectable halide segregation. We explain this intriguing phenomenon by the contributions of two main factors to the change of the system free energy, namely the thermal entropy and lattice strain that are increased during the heating and colling processes, respectively.

Following the same procedure as reported previously[41], we synthesize cuboid $CsPbBr_{1.2}I_{1.8}$ NCs with an average edge length of ~23 nm (see Methods). As depicted in Fig. 1a, we then place a high-density film of $CsPbBr_{1.2}I_{1.8}$ NCs in between two electrodes on top of an ITO substrate, which stays at room temperature of ~23-24 °C without any current (0 mA, Fig. 1b). When an external voltage of 6 V is applied to the ITO substrate, its temperature first increases quickly to ~32 °C within 10 s and then rises further slowly to ~37 °C after ~600 s of current heating (400 mA, Fig. 1b). Compared to the traditional method of heating by a hot plate[42-44], the current heating of ITO substrate can achieve temperature change of the deposited $CsPbBr_{1.2}I_{1.8}$ NCs very promptly, thus allowing us to explore how a dynamic variation of the thermal environment affects the phase-segregation process. To this end, we use an immersion-oil objective to focus the 405 nm output beam from a 5 MHz picosecond diode laser onto the sample (Fig. 1a). Optical signals of $CsPbBr_{1.2}I_{1.8}$ NCs excited at a power density of ~50 W/cm$^2$ are collected by the same objective and sent through a spectrometer to the CCD camera for photoluminescence (PL) spectral measurement (see Methods).

At room temperature without any current heating, the phase segregation of $CsPbBr_{1.2}I_{1.8}$ NCs completes within ~10 min of light illumination, which is evidenced by the optical measurements in the top panel of Fig. 1c. During this process, the PL peak wavelength shifts from the initial ~626 nm to the final ~505 nm, indicating the formation of $CsPbBr_3$ NCs as a result of halide segregation[41]. In addition to the new PL peak, a residual one can also be seen at ~626 nm, which originates from those $CsPbBr_{1.2}I_{1.8}$ NCs located at the edge of the laser spot without receiving enough power density to trigger phase segregation[26,37]. The bottom panel of Fig. 1c shows the recorded PL spectra when the sample is then being left in the dark while the laser beam is unblocked for 1 s only at several time points. Compared to the light-illuminating duration of ~10 min to fully induce the phase segregation, a much longer time of ~120 min is needed for the recovery of the mixed phase.

In the next experiment, we apply an external voltage to the ITO substrate to induce a current flowing of 400 mA in the dark, and a stable temperature of ~35 °C can be achieved after 300 s without triggering phase segregation in the $CsPbBr_{1.2}I_{1.8}$ NCs (Supplementary Fig. 1). The PL spectrum measured right after laser excitation is centered at a shorter wavelength of ~617 nm with a broadened linewidth (Fig. 1d) relative to the one of ~626 nm acquired at room temperature (Fig. 1c). These heating-induced changes can be jointly contributed by the enlarged crystal lattice and the enhanced exciton-phonon coupling commonly observed in perovskite materials[45,46]. Compared to the time duration of ~10 min for the completion of phase segregation at room temperature, a much shorter time of ~90 s is now needed at ~35 °C due to the more active diffusion kinetics of halide anions[17,42,47]. After the above treatment, the sample film is being left in the dark still at ~35 °C while the PL spectra are measured at several time points in the bottom panel of Fig. 1d to monitor the reversal process. PL intensity of the mixed-phase peak increases and gets to a stable value after ~25 min, significantly shorter than the room-temperature time of ~120 min due to the enhanced entropic effect favoring a faster halide mixing in the dark[48].

Consistent with literature[42,49,50], the light-induced phase segregation observed here proceeds at a much faster rate than those of the dark-recovered phase-remixing process at both room temperature and ~35 °C. The above observation implies that the driving force provided by light illumination is dominant over the entropic driving force, and phase segregation of $CsPbBr_{1.2}I_{1.8}$

NCs is always preferred at static temperatures. However, a critical point completely neglected in previous studies is that the system entropy can respond instantly to the temperature change, while ion migration requires a much longer time period in the perovskites[17,51,52]. This motivates us to introduce the tuning knob of dynamic temperature change to our perovskite system of $CsPbBr_{1.2}I_{1.8}$ NCs, thinking that a sudden heating or cooling operation might shift the balance between the forward phase-segregation and the reverse phase-remixing processes.

To confirm this hypothesis, we monitor the time-dependent PL spectra of $CsPbBr_{1.2}I_{1.8}$ NCs during two separate experiments, where we show the phase segregation can be suppressed by either the application of heating to the ITO substrate or the removal of heating from it. For the first experiment, as shown in the bottom panel of Fig. 2a, the segregated phase is first induced by ~300 s of light illumination at room temperature. Then right after a current heating of 400 mA is applied to the ITO substrate, the segregated phase completely shifts back to the mixed phase within ~4 s while the $CsPbBr_{1.2}I_{1.8}$ NCs are still being illuminated (top panel, Fig. 2a). The corresponding PL spectra extracted at several time points in Fig. 2b demonstrates the fast phase-remixing process, during which the sample temperature increases from ~24-31°C within 4 s since the start of current heating (Fig. 2c). We then perform a second experiment by first inducing phase segregation in another film location of $CsPbBr_{1.2}I_{1.8}$ NCs after ~90 s of continuous light illumination under the current heating at 400 mA (bottom panel, Fig. 2d). As can be seen from the time-dependent spectral image in the top panel of Fig. 2d, the segregated phase is converted back to the mixed phase within ~8 s after the current heating has been removed. The corresponding PL spectra extracted at several time points are plotted in Fig. 2e for this phase-remixing process, during which a decrease of the sample temperature from ~35-27 °C can be monitored in Fig. 2f. In Supplementary Figs. 2,3, we present quite similar results obtained at other current values of 300 and 500 mA, while the 400 mA current is employed in our experiment as the result of a trade-off between fast heating/cooling and moderate impact of the thermal accumulation.

We note that despite the exciting observations from Fig. 2 that a sudden start (stop) of current heating can restore the mixed phase within ~4 s (~8 s), the segregated phase would be resumed again when the CsPbBr$_{1.2}$I$_{1.8}$ NCs are exposed to further light illumination beyond this transition point (Supplementary Fig. 4). Based on the facts that the mixed phase can be restored temporarily by either increasing or decreasing the temperature, we speculate a well-designed procedure of periodic heating and colling could help to maintain the mixed phase, i.e., heating or cooling the sample at a timing just before the phase starts to segregate. We achieve this by regulating the time width ($t$) of a single 400 mA pulse within a given time period ($T$). The time-dependent current curves with five different $t/T$ combinations are plotted in Fig. 3a, together with the direct-current (DC) one where the 400 mA current is flowing all the time across the ITO substrate.

After 180 s of light illumination with different $t$/T combinations of the applied current, it can be seen from the PL spectra in Fig. 3b that severe phase segregation occurs in the DC and 0.1/1 s cases, while it is much milder under the conditions of 1/10, 2/10 and 5/10 s. Meanwhile, there is no sign of phase segregation in the $t/T$ combination of 0.5/5 s, which can be further revealed from the time-dependent PL spectral revolution in Fig. 3c. As shown in Fig. 3d, the extracted PL intensities remain stable during the 180 s measurement except the occurrence of small variations caused by periodic heating and cooling cycles. In contrast, the sample under DC current exhibits a drastic trend of PL decaying caused by phase segregation at the first place and then by thermal quenching of the photo-generated excitons. In fact, even for the 0.5/5 s case that allows 4.5 s of sample cooling within every 5 s, local heat can still be built up to cause a slight blue shift in the PL peak and a reduced PL intensity after 180 s of simultaneous light illumination and current heating (Supplementary Fig. 5). Such thermal instability can be circumvented in practice such as by improving the ventilation condition and adopting a heat sink, which are common procedures in the operations of perovskite optoelectronic devices[53].

We emphasize that for the optical experiments performed in Figs. 2,3, light illumination is always supplied to the $CsPbBr_{1.2}I_{1.8}$ NCs during the sudden change of sample temperature, highlighting that the segregated phase can be reversed to the mixed phase even under such a harsh condition. Now if the segregated $CsPbBr_{1.2}I_{1.8}$ NCs are being left in the dark instead, a sudden temperature change would also promote a fast appearance of the mixed phase (Supplementary Fig. 6). Thereafter, no further phase segregation occurs due to the lack of the driving force provided by light illumination. These results allow us to decouple the effect of light illumination with that of sample heating/cooling, and then the main scientific question is how the latter operation could accelerate the phase-remixing process in the dark. Since previous experimental[12,26,42,43,48] and theoretical[23-25] studies of mixed-halide perovskites were mostly done at static temperatures, a new mechanism should be at play in our case of the dynamic temperature response of segregated $CsPbBr_{1.2}I_{1.8}$ NCs towards phase remixing.

Taking all our experimental findings together, we propose a mechanism that takes into account several unique properties of mixed-halide perovskites in the form of NCs. Under light illumination, the excited charge carriers in a single $CsPbBr_{1.2}I_{1.8}$ NC could be captured by surface traps to create local electric fields[22,32] that are capable of breaking the lead-halide bonds [41]. The freed iodide and bromide anions are mobile and have the tendency of refilling the vacancies left in the NC under the entropic driving force[23,25]. The respective vacancy-refilling rates of iodide and bromide anions should be smaller and larger than those of the opposite bond-breaking reactions. The above judgement is based on the fact that the Pb-Br bond is associated with a larger binding energy than that of the Pb-I bond[41,54], and is firmly corroborated by the eventual formation of a single CsPbBr$_3$ NC after phase segregation. Considering the slow ion migration process in perovskite materials[17,51,52], it is surprising to see in our experiment that the internal vacancies of a single CsPbBr$_3$ NC can be occupied by iodide anions within several seconds after a sudden temperature change (Fig. 2a,d). This contradiction can be reconciled by assuming that the freed iodide anions are located in close proximity to the CsPbBr$_3$ NC with a Gaussian-like distribution, as depicted in Fig. 4a. This spatial distribution is a result of the competition of the attraction from internal vacancies to keep iodide anions piling up on the CsPbBr$_3$ NC, with the repulsion among iodide anions to push some of them to nearby regions.

For the coupled system in Fig. 4a where a single CsPbBr$_3$ NC is surrounded by freed iodide anions, the phase-remixing process can be conveniently assumed to start from a vacancy-filling event involving an iodide anion located at the center of the spatial distribution. A neighbouring iodide anion will take the void position left by its predecessor due to the vanished Coulomb repulsion, thus bringing about another round of vacancy-filling and anion-shifting motions until a single CsPbBr$_{1.2}$I$_{1.8}$ NC is completely restored after tens of minutes (bottom panels, Fig. 1c,d). As shown in Fig. 4b for the segregated phase, a sudden operation of sample heating with an elevated entropic driving force promotes vacancy fillings by more iodide anions as compared to the mild situation in Fig. 4a. The greatly-reduced Coulombe repulsion triggers an avalanched movement of iodide anions towards the central region and speeds up the collapse of their spatial distribution, so that phase remixing can be reached in just a few seconds (top panel, Fig. 2a). Bearing the above picture in mind to depict the phase-remixing process after a sudden temperature increase, it is counter-intuitive to understand why the same thing happens after a sudden temperature decrease that seems to attenuate the entropic driving force of remixing.

Due to the size difference between bromide (~196 pm) and iodide (~220 pm) anions[55,56], the crystal lattice of a single $CsPbBr_{1.2}I_{1.8}$ NC hosts extensive local strains, which can be effectively attenuated from the single $CsPbBr_3$ NC depicted in Fig. 4a owing to a total loss of the iodide elements. Being exposed to a sudden decrease of temperature, as shown in Fig. 4c, the $CsPbBr_3$ NC experiences a compressive strain in the contracted lattice. To mitigate such strain, the iodide anions in the distribution center at the surface of the NC are activated to fill the internal vacancies. This analysis is consistent with the literature scenarios, where the grain boundary[57,58] and polaronic region[19,20] with abundant strains are preferred as the migration destinations of iodide anions to reduce the system free energy. Here, the strain-relieving driving force is larger than just compensating the reduced entropy-mixing force, as can be deduced from the fact that a single $CsPbBr_{1.2}I_{1.8}$ NC is recovered within several seconds after a sudden temperature decrease. Now we can have another look at Fig. 4b for the temperature-increasing case, where tensile strains are induced by lattice expansion of the single $CsPbBr_3$ NC. Under this situation, the pairing of iodide anions and internal vacancies should not increase the tensile strains any further if not decreasing them, so that the enhanced entropic driving force serves as the main factor for the fast recovery of the mixed phase.

To summarize, we demonstrate in $CsPbBr_{1.2}I_{1.8}$ NCs that a sudden temperature change of ~10 °C in the dark can reduce the transition time from segregated to mixed phases by more than two orders of magnitude as compared to that measured at static temperatures. Even under laser excitation at a high power density of ~50 W/cm$^2$, this mixed phase can still be temporarily obtained by a sudden temperature change, and it can be further kept permanently when the $CsPbBr_{1.2}I_{1.8}$ NCs are periodically heated with a well-designed time interval. Of special note is that besides the $CsPbBr_{1.2}I_{1.8}$ composition focused here, the temperature-modulation strategy is also applicable for mixed-halide NCs with other bromide-to-iodide ratios (see

Supplementary Fig. 7). To explain all the experimental findings, we propose that the iodide anions are released from the broken Pb-I bonds of a single CsPbBr$_{1.2}$I$_{1.8}$ NC under light illumination and migrate to the surface to form a Gaussian-like spatial distribution around the converted CsPbBr$_3$ NC. This spatial distribution of surface-accumulated iodide anions is kept by the balancing forces of their mutual repulsion and the attraction from internal vacancies, which can be interrupted under an appropriate driving force such as a sudden change of the sample temperature. This temperature change increases either the entropic driving force in the case of heating or the lattice strain in the case of cooling for the segregated phase. To minimize system free energy, the internal vacancies of a single CsPbBr$_3$ NC start to be refilled in an avalanched fashion to promote a fast and complete restoration of the mixed phase. After the dynamic parameter of temperature change has been introduced by our current work, we expect wider applicability of our approach in mixed halide perovskites of various compositions and morphologies to solve the long-standing phase-segregation problem during practical operations of various optoelectronic devices.

## METHODS

**Sample preparations.** To prepare the Cs-oleate precursor, 0.814 g Cs$_2$CO$_3$, 2.5 mL oleic acid and 40 mL octadecene were loaded into a 100 mL three-neck flask. After being degassed and dried for 10 min in vacuum, the solution was heated first at 120 °C for 1 h and then at 150 °C for 2 h under the N$_2$ gas to completely dissolve the Cs$_2$CO$_3$. To prepare the CsPbBr$_{1.2}$I$_{1.8}$ NCs, 0.105 g PbI$_2$, 0.052 g PbBr$_2$, 1.0 mL oleic acid, 1.0 mL oleylamine and 10 mL octadecene were loaded into a 25 mL 3-neck flask. After the solution had been degassed in vacuum and dried for 1h at 120 °C, the flask was connected to the N$_2$ gas. The solution temperature was then raised to 160 °C for 10 min, after which 1.0 mL of the Cs-oleate solution preheated to 120 °C was quickly injected. The reaction was stopped with ice bath after 5 s, and the resulting solution was centrifuged for 20 min at 5000 rpm. The precipitate was re-dispersed in 10.0 mL hexane and centrifuged again for 20 min at 5000 rpm to get the final product of the $CsPbBr_{1.2}I_{1.8}$ NCs in the supernatant. For the synthesis of $CsPbBr_{2.1}I_{0.9}$ NCs, a quite similar procedure was adopted except that the amounts of $PbI_2$ and $PbBr_2$ were adjusted to 0.051 and 0.095 g, respectively. To prepare a high-density film for the optical measurements, one solution drop of the as-synthesized $CsPbBr_{1.2}I_{1.8}$ or $CsPbBr_{2.1}I_{0.9}$ NCs was spin-coated between two electrodes separated by 14 mm on top of an ITO substrate. Each of the deposited electrodes had a width of 2 mm and consisted of the Cr and Au layers (from bottom to top) with the thicknesses of 5 and 70 nm, respectively.

**Experimental measurements.** The sample substrate was attached to a home-built confocal scanning optical microscope, where the $CsPbBr_{1.2}I_{1.8}$ or $CsPbBr_{2.1}I_{0.9}$ NCs were excited at room temperature by a 405 nm picosecond diode laser operated at the repetition rate of 5 MHz. After passing through an immersion-oil objective with the numerical aperture of ~1.4, the laser beam was focused onto the sample substrate with a spot size of ~500 nm and the associated power density of ~50 W/cm$^2$. Optical emission from the $CsPbBr_{1.2}I_{1.8}$ or $CsPbBr_{2.1}I_{0.9}$ NCs were collected by the same objective and sent through a 0.5 m spectrometer to a CCD camera for the PL spectral measurement with an integration time of 0.5 s. For the heating operation, a bias voltage was applied to the two electrodes of the ITO substrate by a digital source meter (Keithley2636B). The surface temperature of the NC film was monitored by a non-contact thermometer, with the obtained data being recorded and processed by the TH11S-B software.

**Data availability**

The data supporting the findings of this study are available from the corresponding authors upon request.

**Acknowledgements**

S.T. acknowledges funding supports from NWO START-UP (No. 740.018.024) and VIDI (No. VI.Vidi.213.091) from the Netherlands. X.W. acknowledges funding supports from the National Basic Research Program of China (Nos. 2021YFA1400803 and 2019YFA0308704), the National Natural Science Foundation of China (Nos. 62174081 and 61974058), and the Priority Academic Program Development of Jiangsu Higher Education Institutions.

**Author contributions**

X.W., C.Z. and M.X. conceived and designed the experiments., S.F., R.D., Y.J. and S.L. prepared the samples and performed the optical measurements. S.F. and X.W. analyzed the data. X.W. and S.T. developed the theoretical model. X.W., S.T. and S.F. co-wrote the manuscript with the help of C.Z. and M.X.

**Additional information**

The authors declare no competing financial or non-financial interests. Correspondence and requests for materials should be addressed to S.T., M.X. or X.W.

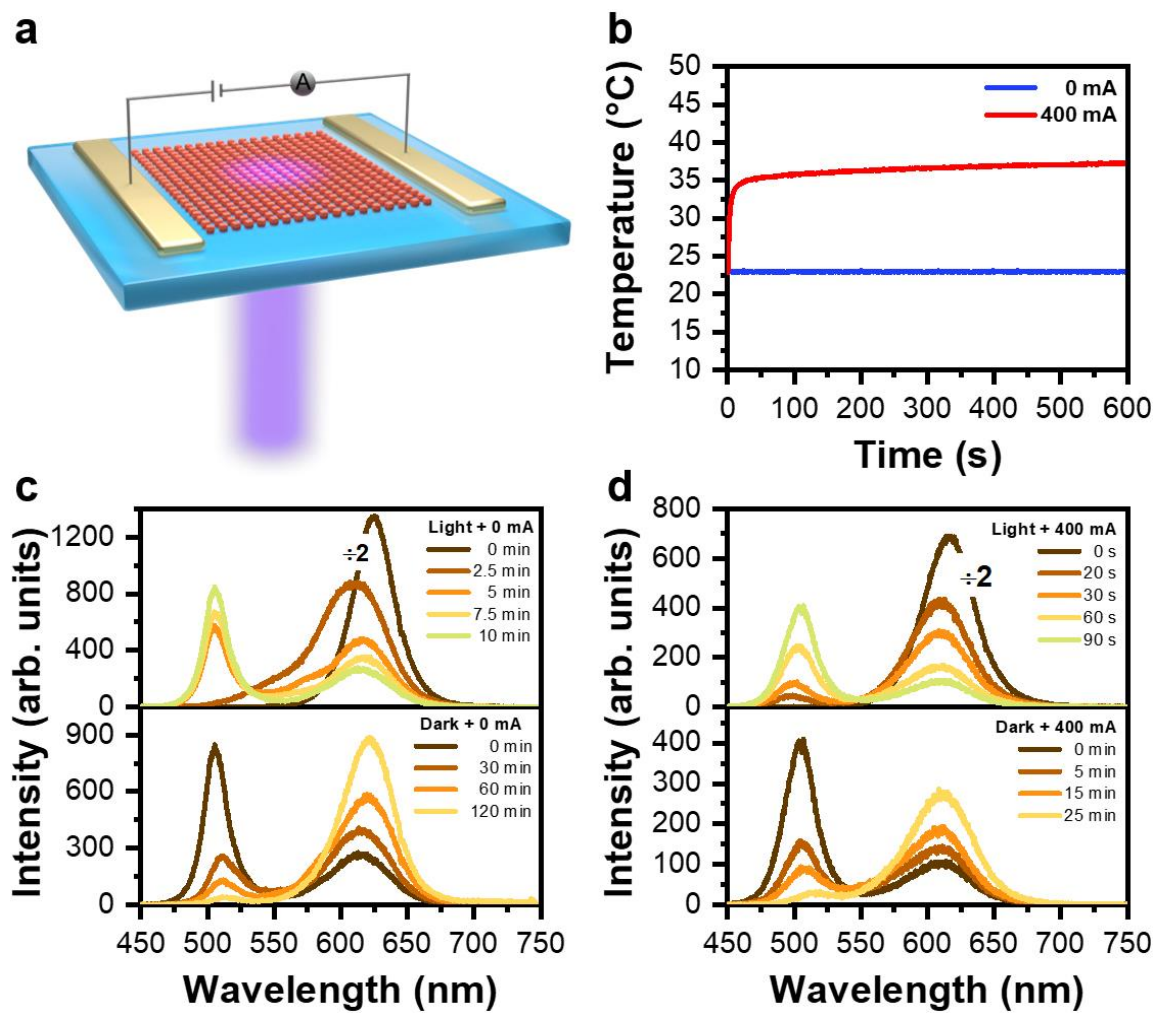

**Figure 1 | Phase-segregation studies of CsPbBr$_{1.2}$I$_{1.8}$ NCs at static temperatures. a,** Experimental setup for the sample heating/cooling operations and the optical studies. **b,** Temperature evolution of CsPbBr$_{1.2}$I$_{1.8}$ NCs after a current flowing of 400 mA is induced in the ITO substrate, as compared to that measured at room temperature (0 mA). PL spectra measured during the light-induced phase-segregation (top) and dark-recovered phase-remixing (bottom) processes of CsPbBr$_{1.2}$I$_{1.8}$ NCs at **c,** room temperature and **d,** ~35 °C.

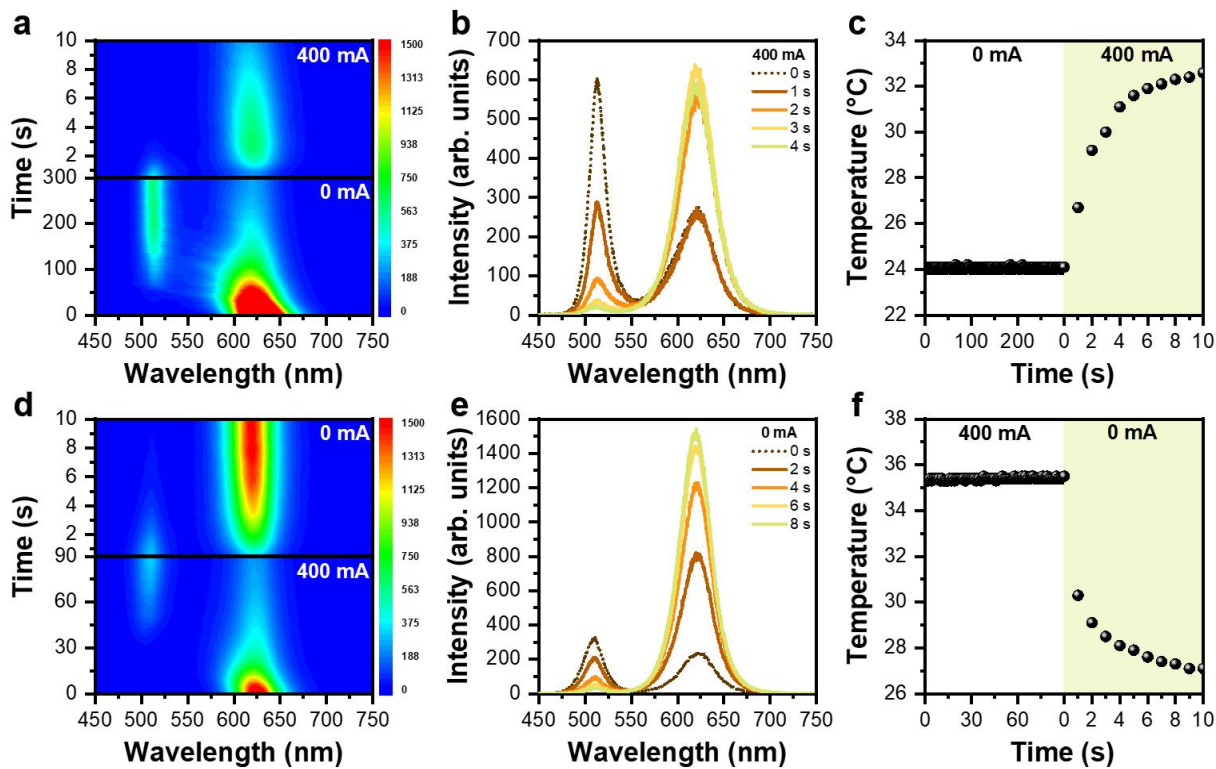

**Figure 2 | Phase-segregation studies of CsPbBr$_{1.2}$I$_{1.8}$ NCs at dynamic temperatures. a,** Time-dependent spectral images measured before (bottom) and after (top) a current flowing of 400 mA is induced in the ITO substrate. **b,** PL spectra extracted from the top panel of **a** at several time points from 0-4 s. **c,** Sample temperatures monitored before and after a current flowing of 400 mA is induced in the ITO substrate. **d,** Time-dependent spectral images measured before (bottom) and after (top) a current flowing of 400 mA is removed from the ITO substrate. **e,** PL spectra extracted from the top panel of **d** at several time points from 0-8 s. **f,** Sample temperatures monitored before and after a current flowing of 400 mA is removed from the ITO substrate.

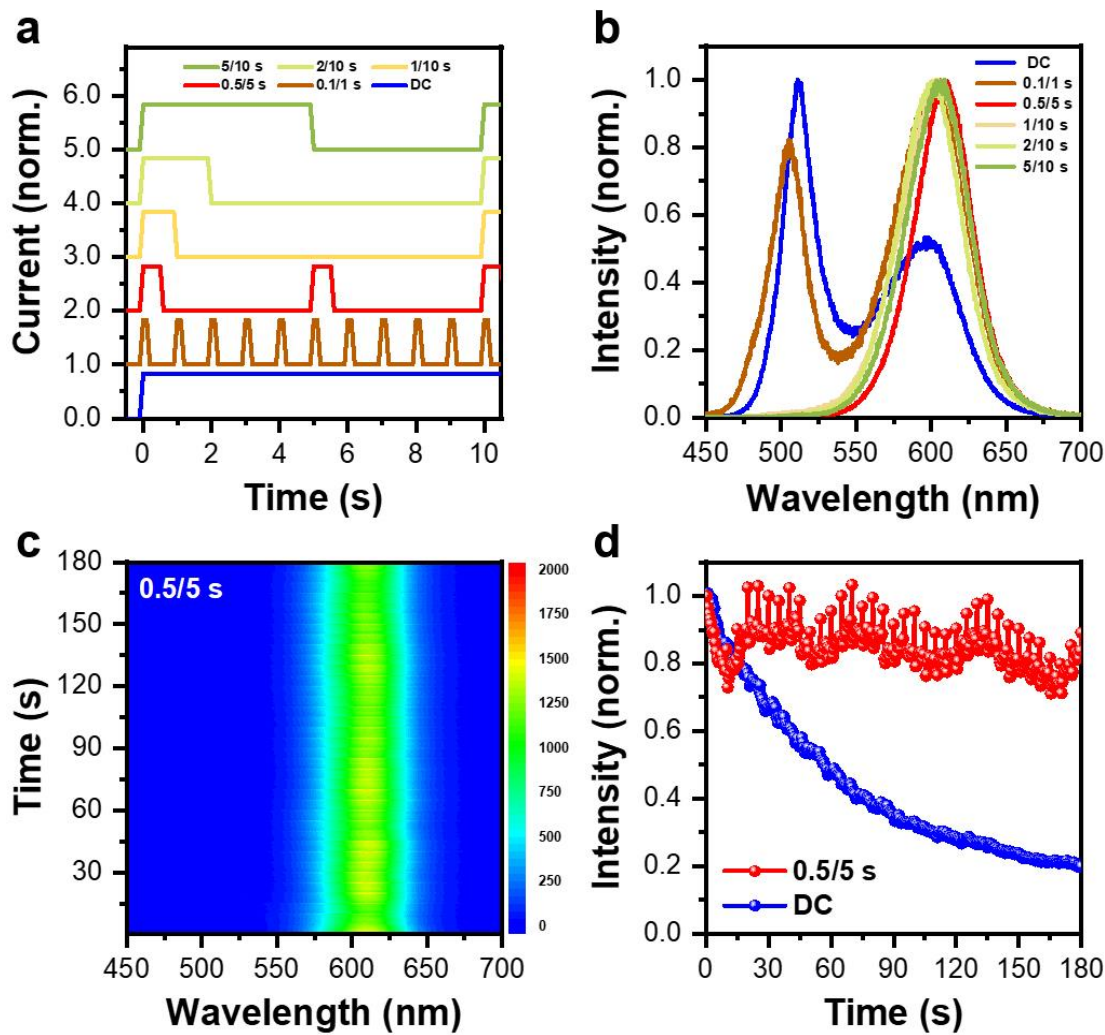

**Figure 3 | Complete suppression of phase segregation in CsPbBr$_{1.2}$I$_{1.8}$ NCs. a,** Different current patterns applied to the ITO substrate, normalized to their maximum amplitudes of 400 mA and offset to each other for clarity. **b,** PL spectra measured for CsPbBr$_{1.2}$I$_{1.8}$ NCs after different current patterns have been applied to the ITO substrate for 180 s. **c,** Time-dependent spectral image measured for CsPbBr$_{1.2}$I$_{1.8}$ NCs with the current pattern of 0.5/5 s. **d,** Mixed-phase PL intensities of CsPbBr$_{1.2}$I$_{1.8}$ NCs plotted as a function of time for the two current patterns of DC and 0.5/5 s, respectively. In the optical measurement with a specific current pattern, a different film position of CsPbBr$_{1.2}$I$_{1.8}$ NCs is selected to acquire the PL spectrum.

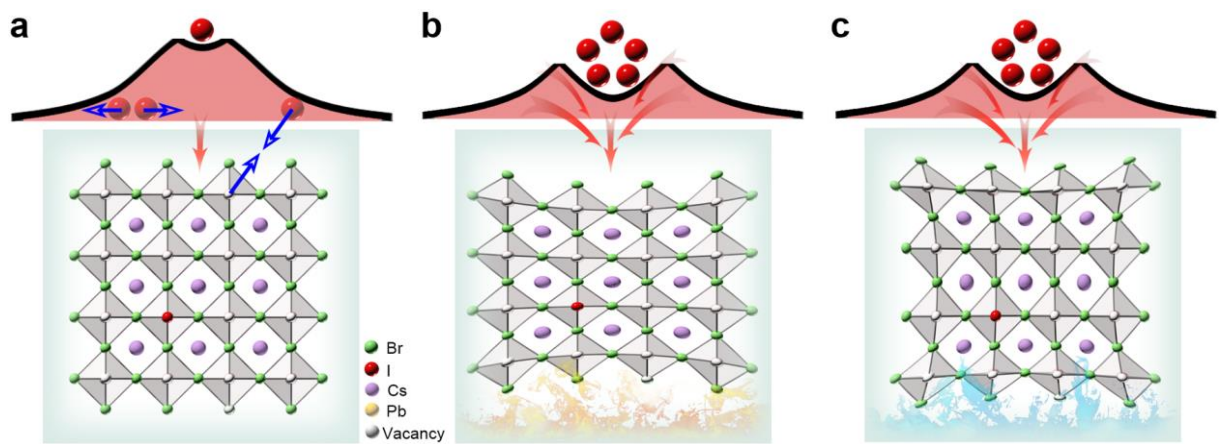

**Figure 4 | Underlying mechanism for heating- and cooling-induced phase remixing in the dark. a,** Spatial distribution of freed iodide anions in close proximity to a single CsPbBr$_3$ NC, with the central one just about to fill the internal vacancy to restore a broken Pb-I bond. This spatial distribution of iodide anions is maintained by their mutual repulsions and the attractions from internal vacancies. **b,** When the CsPbBr$_3$ NC is suddenly heated to have a lattice expansion, more iodide anions in the distribution center would refill the vacancies to cause a fast restoration of the mixed phase. **c,** When thermal heating is suddenly removed from the CsPbBr$_3$ NC to induce a lattice contraction, more iodide anions in the distribution center would refill the vacancies to cause a fast restoration of the mixed phase.

**Supplementary Information**

**Complete Suppression of Phase Segregation in Mixed-Halide Perovskite Nanocrystals under Periodic Heating**

Shengnan Feng[1†], Rentong Duan[1†], Yu Ju[1], Shuyi Li[1], Chunfeng Zhang[1], Shuxia Tao[2*], Min Xiao[1,3*], and Xiaoyong Wang[1*]

[1]*National Laboratory of Solid State Microstructures, School of Physics, and Collaborative Innovation Center of Advanced Microstructures, Nanjing University, Nanjing 210093, China*

[2]*Materials Simulation and Modelling & Center for Computational Energy Research, Department of Applied Physics, Eindhoven University of Technology, 5600 MB Eindhoven, The Netherlands*

[3]*Department of Physics, University of Arkansas, Fayetteville, Arkansas 72701, USA*

[*]Correspondence to S.T. (S.X.Tao@Tue.nl), M.X. (mxiao@uark.edu) or X.W. (wxiaoyong@nju.edu.cn)

[†]These authors contributed equally to this work

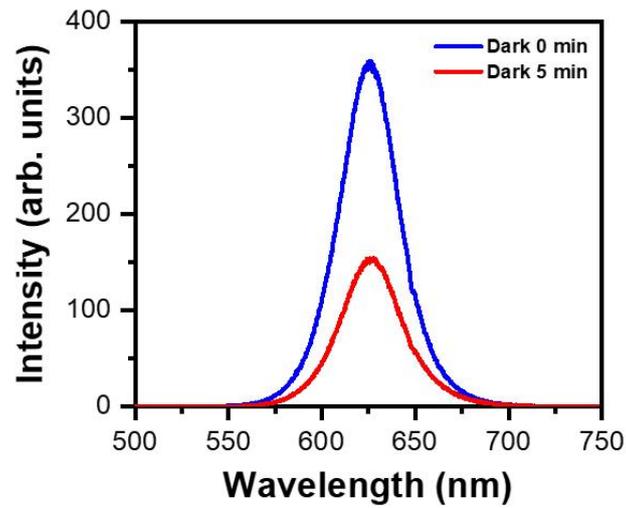

**Supplementary Figure 1.** PL spectra measured for $CsPbBr_{1.2}I_{1.8}$ NCs at the beginning and after 5 min of the current heating at ~400 mA. The $CsPbBr_{1.2}I_{1.8}$ NCs are left in the dark while the excitation laser beam is unblocked only for 1s for the PL spectral measurements.

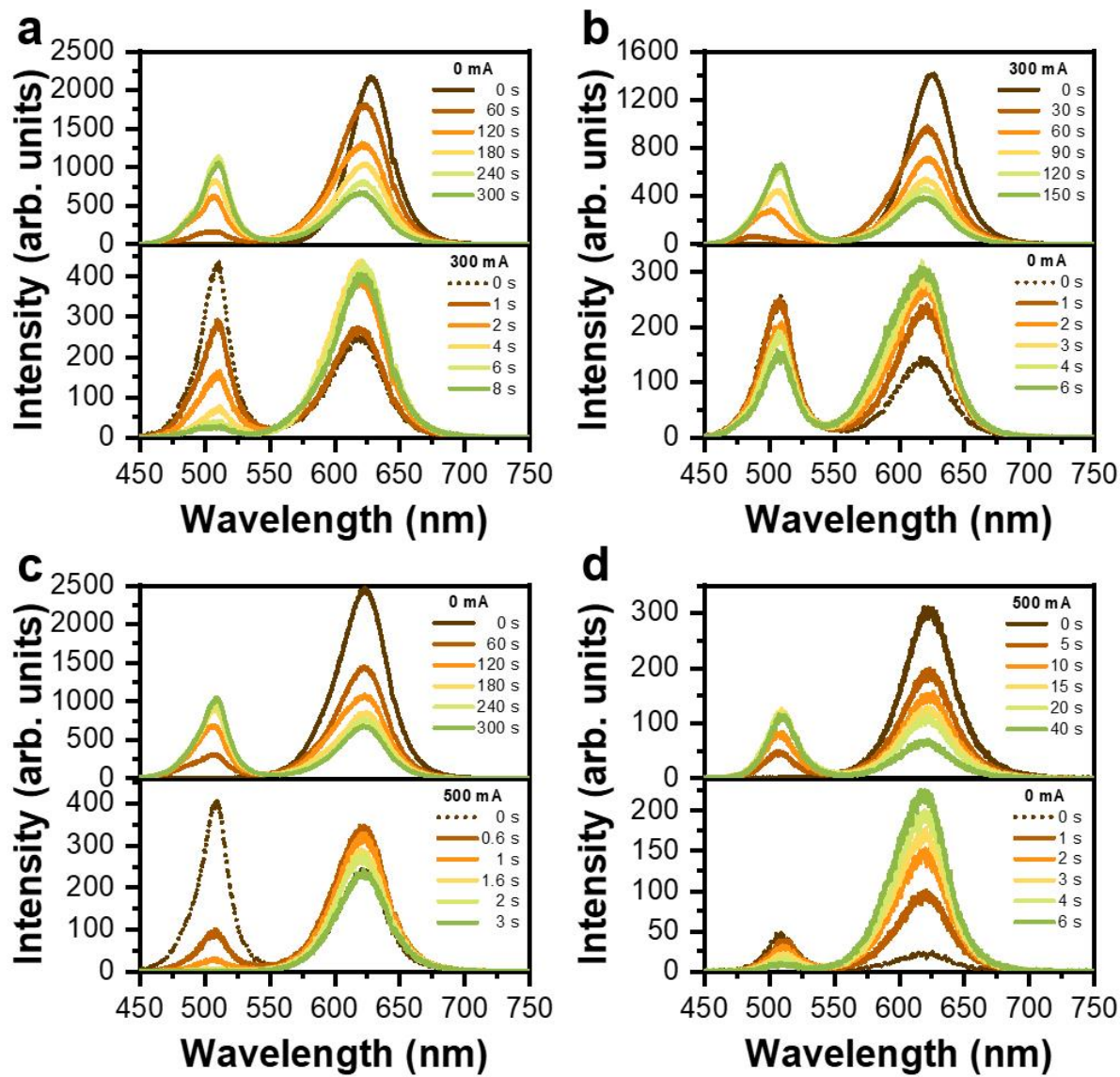

**Supplementary Figure 2. a,** (top) PL spectra measured at several time points during the light-induced phase-segregation process of CsPbBr$_{1.2}$I$_{1.8}$ NCs at room temperature. (bottom) PL spectra measured at several time points after a 300 mA current is applied to the ITO substrate, inducing a fast transition from the segregated to the mixed phases under light illumination. **b,** (top) After a 300 mA current has been applied to the ITO substrate for 300 s to reach a stable temperature in the dark, the PL spectra are measured at several time points during the light-induced phase-segregation process of CsPbBr$_{1.2}$I$_{1.8}$ NCs. (bottom) PL spectra measured at several time points after the 300 mA current is removed from the ITO substrate, inducing a fast transition from the segregated to the mixed phases under light illumination. **c,d,** Similar experimental results to those demonstrated in **a,b**, except that the applied current is changed from 300 to 500 mA.

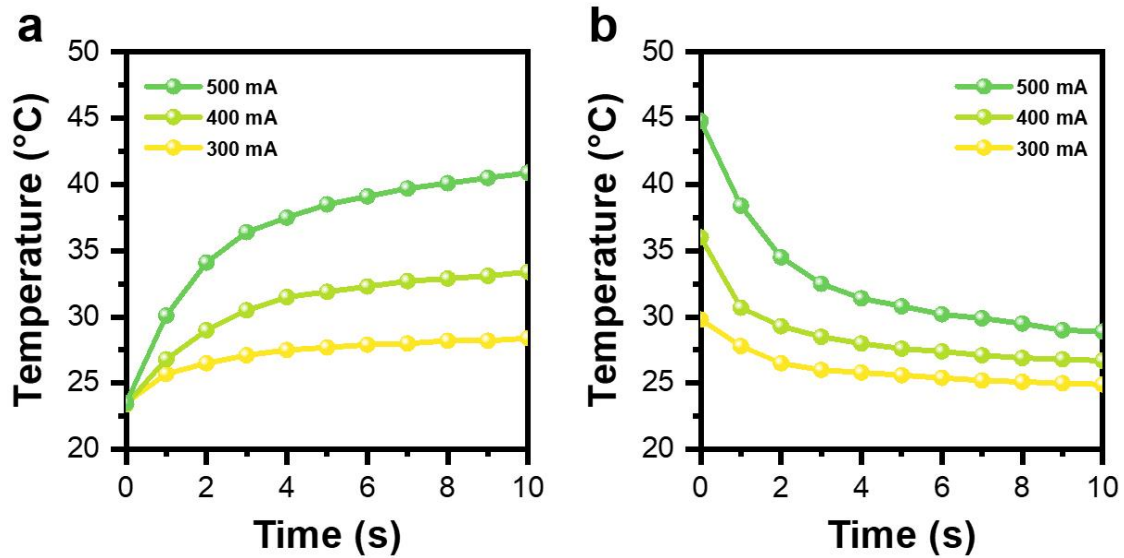

**Supplementary Figure 3. a,** Sample temperatures monitored right after the currents of 300, 400 and 500 mA have been applied to the ITO substrate. **b,** Sample temperatures monitored right after the currents of 300, 400 and 500 mA have been removed from the ITO substrate.

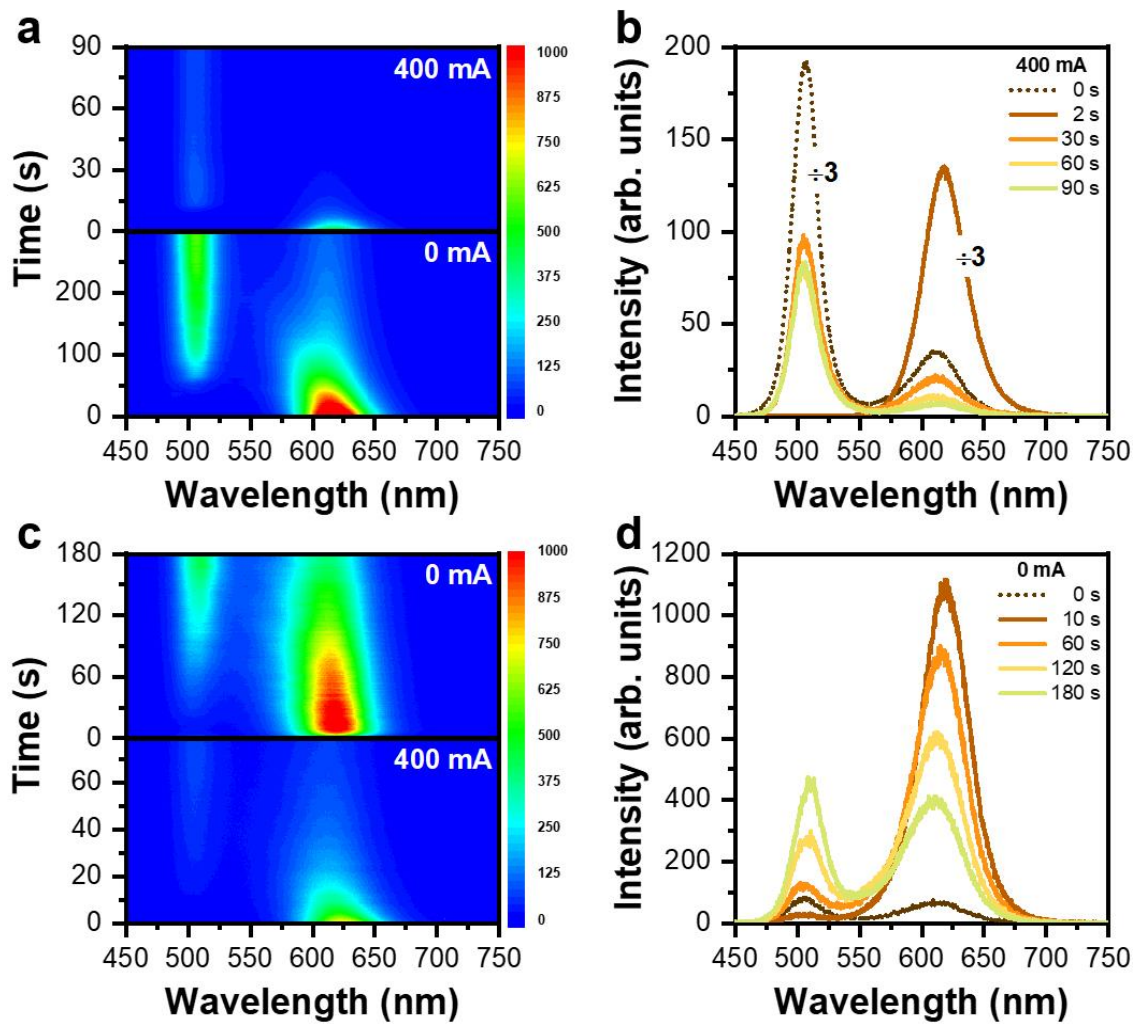

**Supplementary Figure 4. a,** Time-dependent spectral image measured for CsPbBr$_{1.2}$I$_{1.8}$ NCs before (bottom) and after (top) a 400 mA current is applied to the ITO substrate. **b,** PL spectra extracted from the top panel of **a** at several time points between 0-90 s. **c,** Time-dependent spectral image measured for CsPbBr$_{1.2}$I$_{1.8}$ NCs before (bottom) and after (top) a 400 mA current is removed from the ITO substrate. **d,** PL spectra extracted from the top panel of **c** at several time points between 0-180 s.

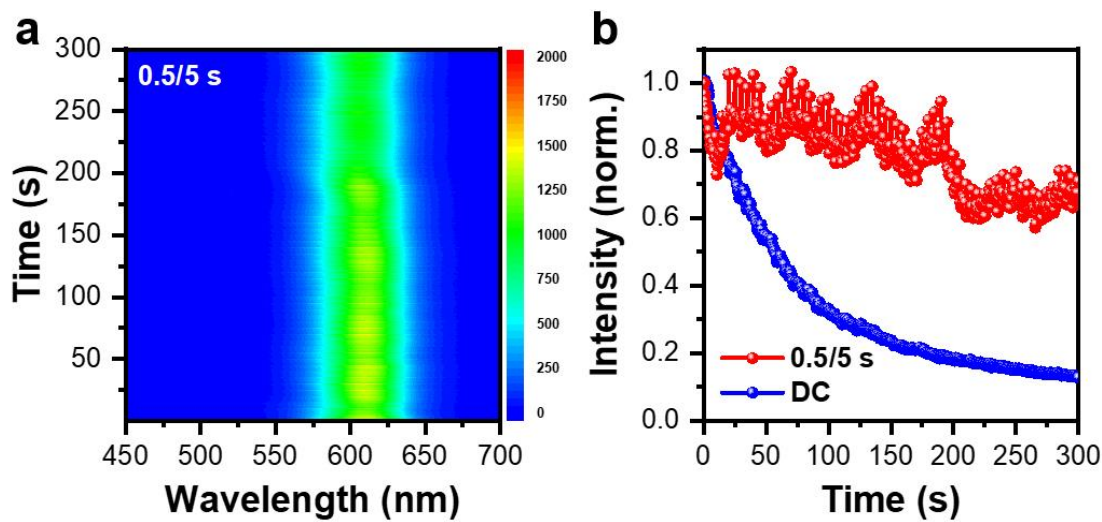

**Supplementary Figure 5. a,** Time-dependent spectral image measured for CsPbBr$_{1.2}$I$_{1.8}$ NCs with the current pattern of 0.5/5 s. **b,** Mixed-phase PL intensities of CsPbBr$_{1.2}$I$_{1.8}$ NCs plotted as a function of time for the two current patterns of DC and 0.5/5 s, respectively. Compared to Fig. 3**c,d** in the main text, the time-dependent spectral image and PL intensities are plotted here at an enlarged time window from 0-300 s.

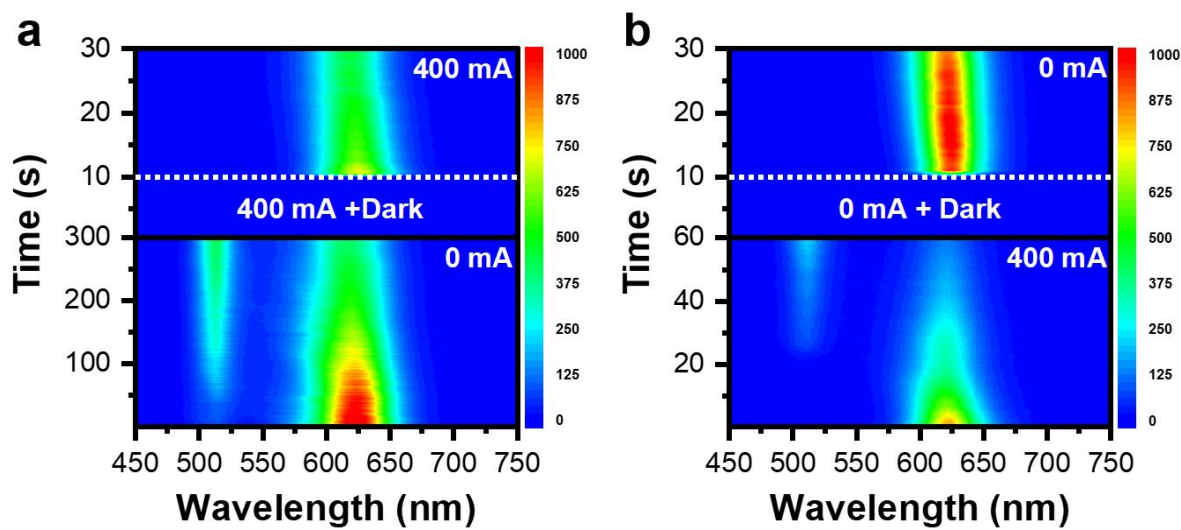

**Supplementary Figure 6. a,** (bottom) Time-dependent spectral image showing light-induced phase segregation of CsPbBr$_{1.2}$I$_{1.8}$ NCs at room temperature. (middle) The laser excitation beam is then blocked with a 400 mA current being applied to the ITO substrate for 10 s. (top) Time-dependent spectral image of the same CsPbBr$_{1.2}$I$_{1.8}$ NCs measured after the laser excitation beam is unblocked again. **b,** (bottom) Time-dependent spectral image showing light-induced phase segregation of CsPbBr$_{1.2}$I$_{1.8}$ NCs with a 400 mA current being applied to the ITO substrate. (middle) The current heating is then removed from the ITO substrate and the laser excitation beam is blocked for 10 s. (top) Time-dependent spectral image of the same CsPbBr$_{1.2}$I$_{1.8}$ NCs measured after the laser excitation beam is unblocked again.

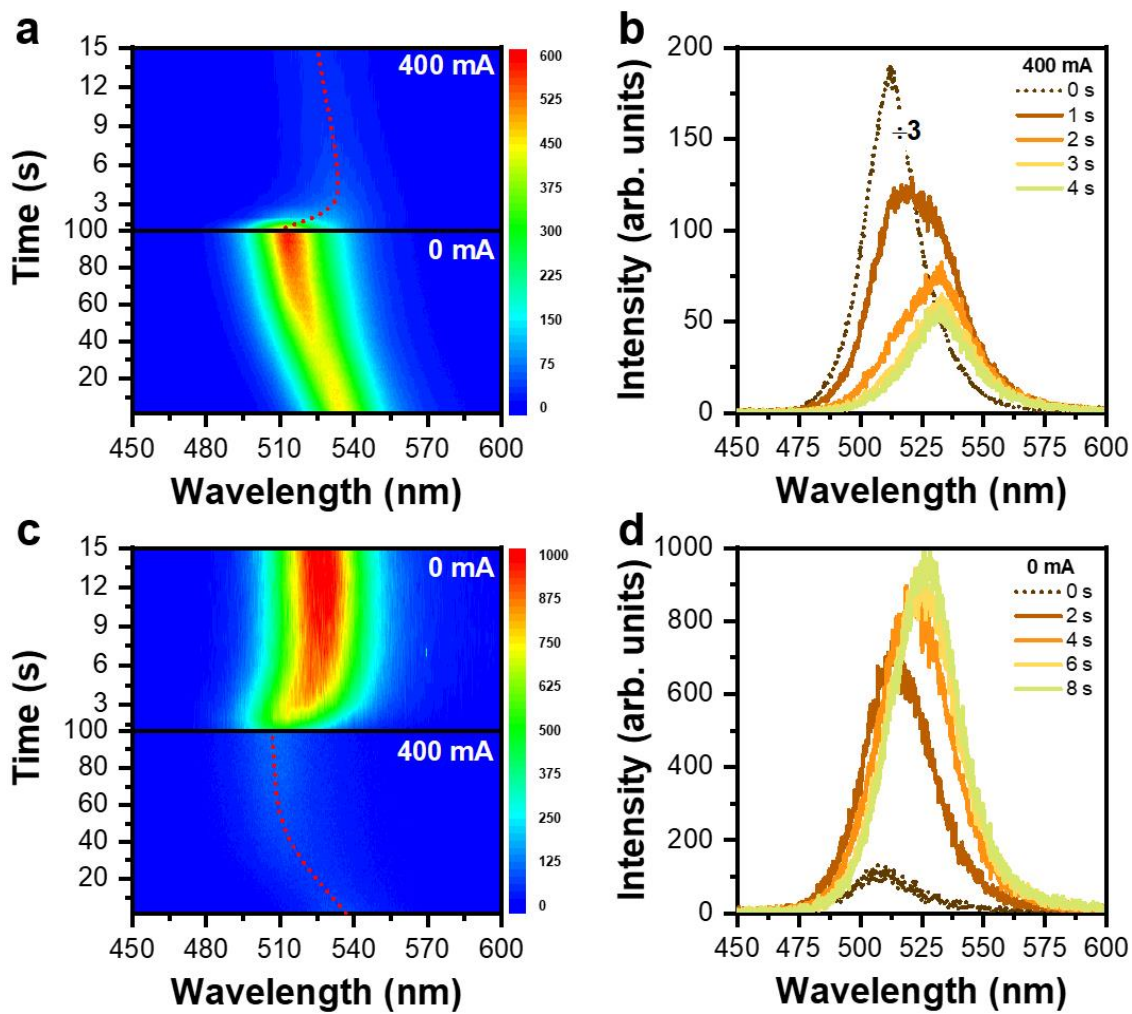

**Supplementary Figure 7. a,** Time-dependent spectral image measured for $CsPbBr_{2.1}I_{0.9}$ NCs before (bottom) and after (top) a 400 mA current is applied to the ITO substrate. **b,** PL spectra extracted from the top panel of **a** at several time points from 0-4 s. **c,** Time-dependent spectral image measured for $CsPbBr_{2.1}I_{0.9}$ NCs before (bottom) and after (top) a 400 mA current is removed from the ITO substrate. **d,** PL spectra extracted from the top panel of **c** at several time points from 0-8 s. In **a** or **c**, the PL intensity of $CsPbBr_{2.1}I_{0.9}$ NCs becomes too weak under the 400 mA current heating, so that a red-dotted line is used to mark the PL peak position as a function of time.